\documentclass[twocolumn,prd,unsortedaddress,superscriptaddress,showpacs,a4paper,nofootinbib]{revtex4}
\usepackage{epsfig}
\def\beq{\begin{equation}}
\def\enq{\end{equation}}
\def\beqa{\begin{eqnarray}}
\def\enqa{\end{eqnarray}}

\def\G3{\lag g^3G^3\rag}

\def\la{\lambda}

\def\si{\sigma}

\def\al{\alpha}
\def\be{\beta}

\def\pkpi{p_{K\pi}}
\def\ppipi{p_{\pi\pi}}
\def\nn{\nonumber}
\newcommand{\rag}{\rangle}
\newcommand{\lag}{\langle}

\begin{document}

\title{\sc Final state hadronic interactions and 
           non-resonant $B^\pm\to K^\pm\pi^+\pi^-$ decays}
\author{I. Bediaga}
\email{bediaga@cbpf.br}
\affiliation{Centro Brasileiro de Pesquisas F\'\i sicas, Rua Xavier 
Sigaud 150, 22290-180 Rio de Janeiro, RJ, Brazil}
\author{D.R. Boito}
\email{dboito@if.usp.br}
\affiliation{Instituto de F\'{\i}sica, Universidade de S\~{a}o Paulo, 
C.P. 66318, 05389-970 S\~{a}o Paulo, SP, Brazil}
\author{G. Guerrer}
\email{guerrer@cbpf.br}
\affiliation{Centro Brasileiro de Pesquisas F\'\i sicas, Rua Xavier 
Sigaud 150, 22290-180 Rio de Janeiro, RJ, Brazil}
\author{F.S.~Navarra}
\email{navarra@if.usp.br}
\affiliation{Instituto de F\'{\i}sica, Universidade de S\~{a}o Paulo, 
C.P. 66318, 05389-970 S\~{a}o Paulo, SP, Brazil}
\author{M.~Nielsen}
\email{mnielsen@if.usp.br}
\affiliation{Instituto de F\'{\i}sica, Universidade de S\~{a}o Paulo, 
C.P. 66318, 05389-970 S\~{a}o Paulo, SP, Brazil}

\begin{abstract}
We evaluate the non-resonant decay amplitude of the process $B^\pm\to K^\pm\pi^+
\pi^-$ using an  approach  based on  final state hadronic 
interactions  described in terms of  meson exchanges. We conclude that this 
mechanism generates inhomogeneities in the Dalitz plot of the $B$ decay.
\end{abstract}

\maketitle

%%%%%%%%%%%%%%%%%%%%%%%%%%%%%%%%%%%%%%%%%%%%%%%%%%%%%%%%%%%%%%%%%%%%%
%
%
%\section{Introduction}
%
%
%%%%%%%%%%%%%%%%%%%%%%%%%%%%%%%%%%%%%%%%%%%%%%%%%%%%%%%%%%%%%%%%%%%%%%%
\section{Introduction}

The amplitude analysis of non-leptonic three body B decays became an
important  tool  to determine the CKM phases \cite{ helen, bediaga, 
silvestrini}, and also to observe CP-asymmetry. Using this method, the  Belle 
and BaBar collaborations have recently \cite{bellek2pi,BaBark2pi} extracted 
a fraction  asymmetry  for  the channel   $B^{\pm} \to K^{\pm} \rho^0$. 
With this kind of analysis  one could also explore the  asymmetry  associated 
with  
%  associated with  the interference between two intermediate states, i.e. 
the interference between two neighbour resonances decaying into the same three 
body final  state.

% through  phase differences  between two charged conjugate decays. 

  The Dalitz plot  analysis needs some  {\it a priori} model, with all possible 
  dynamical components  and  a correct  functional form  to be 
  used to  fit the Dalitz plot distribution. Arbitrary distribution functions 
  can be used to get  a good  fit, but they have no physical meaning. 

  The non-resonant component which, in general, is spread all 
  over the phase space, can mimic other  dynamical components, through 
  the interferences with the  resonances present in the same phase space. 
  This  shadowing phenomenon was observed  in charm three body decays in  
  the E791 experiment \cite{sigma,kappa},  where the overestimated 
  contribution of the non-resonant amplitude replaced, in a wrong way, 
  the contribution (and the very existence) of the scalar mesons 
  $\sigma$ and $\kappa$.

  Belle has proposed, for the amplitude analysis of the process  
   $B^\pm \to K^\pm \pi^+ \pi^-$ \cite{bellek2pi}, a  
   parametrization for the non-resonant amplitude given by: 
\beq
{\cal A}_{nr}(K^\pm\pi^\pm\pi^\mp)=a_1^{nr}e^{-\alpha s_{13}}
e^{i\delta_1^{nr}}+a_2^{nr}e^{-\alpha s_{23}}e^{i\delta_2^{nr}}\;,
\label{anr}
\enq
where $a_1$, $\delta_1$, $a_2$ and  $\delta_2 $ are the fit parameters 
and $s_{13}\equiv M^2(K^\pm\pi^\mp)$ and $s_{23}\equiv M^2(\pi^+\pi^-)$
are the Dalitz variables. Originally the parametrization was also function of 
the variable $s_{12} = M^2 (K^\pm \pi^\pm)$, but the term containing this 
variable turned out to give an insignificant contribution.

The function above could fit the data  
with an acceptable  confidence level around 
1\%. Certainly  this distribution was more useful for the Belle 
analysis than the  usual constant non-resonant distribution. However, it 
has to be employed  with caution.

The use of an empirical parametrization, with a form without  dynamical 
content, may hide  the physical meaning of the Dalitz plot and, even worse, may 
yield  inadequate parameters for the contribution of the resonance amplitudes  
and for the CP asymmetries. This is so because  in the Dalitz distribution 
the parameters are highly correlated among themselves and also  
with the non-resonant amplitude. The parametrization (\ref{anr}) was proposed to 
describe  non-uniformities in the non-resonant three-body decays. Data   
suggest the formation of a $\pi^-$ with momentum predominantly smaller than
the momentum of the $K^+$ and $\pi^+$ in the process  
$B^+ \to K^+ \pi^+ \pi^-$.   How can we understand  this?

In principle there are two different cathegories of dynamical processes 
contributing to the  non-resonant  structures in the Dalitz plot.
One  is a genuine three-body decay that is a consequence 
of the partonic  structure of the weak decay \cite{bgr1,bgr2} and therefore will 
be called parton interaction, or PI. The other comes from  final state  
hadron-hadron   
interactions \cite{la,la2,don,ccs,china,fis} and we shall call it FSI.   
In both approaches  one is forced to make  non-trivial calculation assumptions 
and the  parameters and form factors are not well known. However as 
we are dealing with Dalitz distributions, the comparison 
of these different contributions with the experimental distributions might 
discriminate which dynamical mechanism is more realistic to describe 
data.   

%\section{Qualitative discussion}

In Fig. 1 we show the  diagrams relevant for the 
partonic description of this process.     
Figs. 1a and 1b show the direct and penguin contribution 
to the ``current induced'' processes.  Figs. 1c), 1d) and 1e) show the direct 
(1c)) and 
and penguin (1d) and 1e)) contributions to the ``transition'' processes and 
and Figs. 1f) and 1g)  show the direct and penguin ``annihilation''  processes. 

\begin{figure}[h] 
\centerline{\epsfig{figure=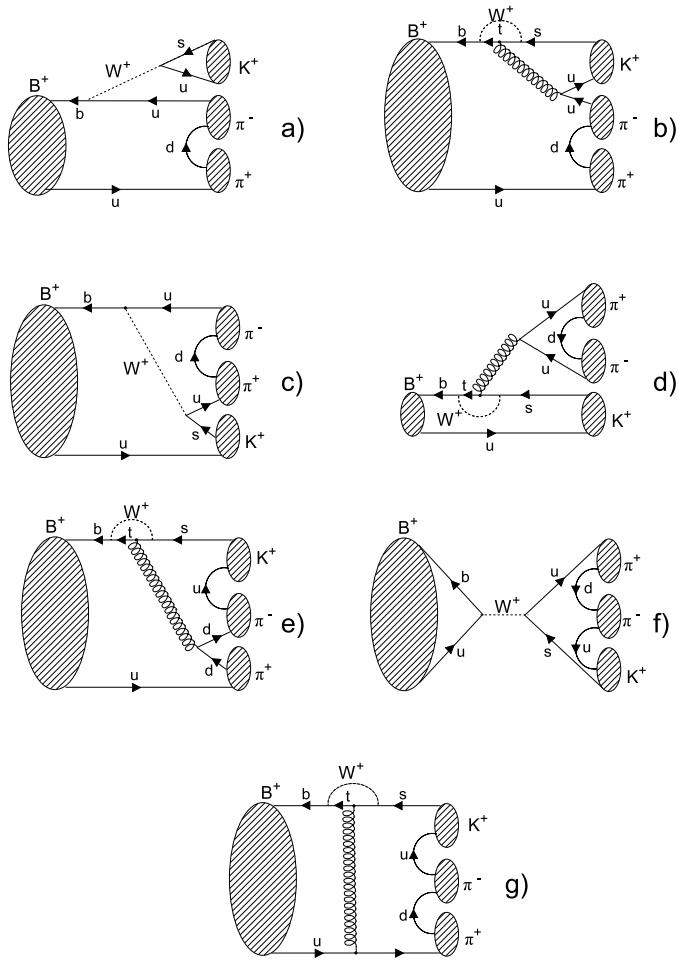,height=140mm}}
\caption{Parton description of  the weak decay $B^{+} \to K^{+} \pi^+ \pi^-$.}
\label{pi1} 
\end{figure}

Looking at diagrams 1a) to 1e) we can see that in the beginning both $\bar{b}$ 
and $u$ 
quarks can  carry a large momentum. When they emit a $W$ or a gluon, the bosons 
can also 
carry a large momentum and therefore, with the exception of the $d$ and 
$\bar{d}$ quarks (which come always from the vacuum and are soft)  any of the 
quarks in the final state 
may carry a large momentum. This large momentum is then transferred to the 
final mesons. 
The final momentum distribution of the three mesons will be eventually  
non-uniform but
there is no reason for producing a softer $\pi^-$. The ``entanglement'' of the  
$\bar{b}$,  $u$ and bosonic lines will distribute more or less democratically 
the initial 
high momenta.  In sharp contrast to  this situation are diagrams 1f) and 1g), 
where after 
the boson emission and absorption the high momenta are {\it only} with the  
$u$  and $\bar{s}$ quarks. In the middle of the diagrams we see the $\pi^-$ 
made of $u$ and $d$ quarks pairs taken from the vacuum. This  $\pi^-$ will be 
comparatively softer than the negative pions of diagrams 1a) - 1e).

\begin{figure}[h] 
\centerline{\epsfig{figure=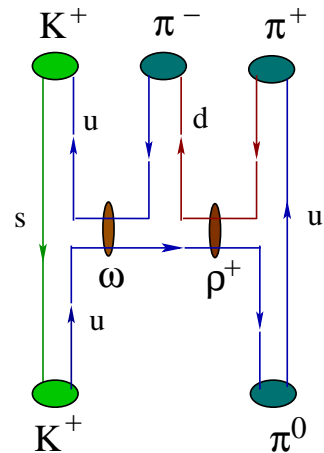,height=60mm}}
\caption{Born diagram for $K^+\pi^0\to K^+\pi^+\pi^-$ final state 
hadronic interaction.}
\label{fsi1} 
\end{figure}

When the final three mesons are produced through  final state interactions  
we can also understand the softer $\pi^-$ 
in a very simple way: first the $B^+$ decays into $K^+$ and $\pi^0$ and then 
these interact creating $K^+$, $\pi^+$ and $\pi^-$. The first two mesons carry 
the valence quarks of the more energetic intermediate  $K^+$ and  $\pi^0$ and 
therefore will have higher 
momenta. The $\pi^-$ is produced with quarks taken from  the vacuum and 
therefore is softer. 
This is illustrated  in Fig.\ref{fsi1}. Thus the appearance of a comparatively
soft $\pi^-$ is nothing 
but the manifestation of the ``leading particle effect'', so frequently 
observed in other processes in hadron physics.

To the best of our  knowledge, this kind of process has been first observed in 
a Dalitz plot by the Aachen-Berlin-CERN Collaboration in  the  reaction  
$\pi^- p \to \pi^+ \pi^0  n $ 
\cite{aachen} and also in $\pi^- p \to K^0 K^0 n $ \cite{brauti}.
Shortly after the experimental observation it was theoretically understood in  
\cite{kajantie}. In that paper it was explained in terms of 
double Regge graphs. This  mechanism implies the  production of a soft pion  
in the central 
region,  responsible for a  bump at the corner of the  Dalitz plot, hence the 
name {\it cornering effect}.  

The analysis  performed in \cite{cca,cy}, suggests that 
in the chiral limit approximation, i.e. when $m_B^2 \, >> \, m_K^2$, diagram 
1f) vanishes.
It  also suggests that the most important process is the one shown in Fig. 1g), 
which has a 
striking similarity with the diagram of Fig. \ref{fsi1}. Both have the same 
final quark flow and  final meson formation.

From the above qualitative discussion and from figures 1g) and \ref{fsi1} 
we conclude that in the decay 
$B^{+} \to K^{+} \pi^+ \pi^-$ we expect
to see a softer $\pi^-$ which will yield a non-uniform Dalitz distribution. This 
effect comes both 
from the partonic weak decay and from final state hadronic interactions and has,
 in both cases, 
the same physical origin:  valence quarks form hard mesons in the final state 
while softer mesons are produced from the vacuum. Neglecting the flavor 
change ($\bar{b} \, \rightarrow \, \bar{s}$) 
in Fig. 1g), we can say that in both cases, PI and FSI, we have a leading 
particle effect.

In what follows we will implement these ideas in a more quantitative way. In view 
of  the  uncertainties in the partonic description, specially the use of the 
factorization hypothesis,  we will develop here 
only  the  hadronic final state interaction  approach, for which the relevant 
lagrangian densities and form factors are better known from a large body of 
phenomenological 
studies at low and intermediate energies. Therefore we shall suppose that the 
non-resonant three-body decay $B^\pm\to  K^\pm\pi^+\pi^-$ proceeds through a 
hadronic scattering between  two intermediate mesons $K^\pm$ and $\pi^0$ in the 
two-body decay $B^\pm\to K^\pm\pi^0$. 

\section{Hadronic final state interactions}

Hadronic final state interactions have a long story and follow  
two different approaches: Regge models and meson exchange models. 
The former have been considered in many works \cite{la,don} whereas the latter 
were
discussed in \cite{ccs,china,fis}. Regge theory is formulated in the high 
energy limit and in 
it all the amplitudes respect unitarity constraints. The interactions are 
represented by 
reggeon exchanges, the Pomeron being usually the most important one. As one 
moves to the low 
energy region, secondary reggeons become important, additional assumptions 
have to be made and 
the theory looses predictive power. It is not clear when the energy starts to 
be ``high'' and 
Regge theory has been applied to energies of the order of a few GeV.  Meson 
exchange models 
describe hadronic interactions fairly well at intermediate energies (a few 
hundreds  MeV). They 
contain  uncertainties associated with higher order diagrams, multiple exchange 
terms, coupling 
constants and the spatial extension of hadrons. These uncertainties are 
translated into form 
factors, which, in turn, can be either calculated (e.g. with QCD sum rules 
\cite{qcdsr}) or simply 
parametrized and fitted to data. A positive aspect of meson exchange models is 
that one can use 
effective Lagrangians for them and so enforce chiral symmetry, which is known 
to be very 
important at intermediate energies. On the other hand, in this kind of model, 
the amplitudes are 
not unitary. The lack of unitarity is believed to become a serious problem at 
increasing 
energies. These considerations suggest that with inceasing energies we should 
change the 
dynamical description, going from meson exchange to reggeon exchange, but is 
not clear 
{\it at which energy} this should be done. In this work we shall study final 
state interactions with meson  dynamics.   

 In principle the $B$ meson can decay into 
inumerous hadronic two-body intermediate states which will subsequently 
decay into $  K^+ \pi^+ \pi^-$. These states include: $K^0 \pi^+$, $K^+ \pi^0$, 
$\eta' K^+$, $\eta' K^{*+}$, $\eta K^+$, $\eta K^{*+}$, $\omega K^+$, 
$\omega K^{*+}$, $a_0^+ K^0$, $a_0^0 K^+$, $K^{*0} \pi^+$, $K^{*+} \pi^0$...
These mesons can undergo a two-to-three reaction, exchanging virtual mesons and 
creating  the $\pi^-$. This is illustrated in Fig. \ref{fsi1} for the 
$K^+ \pi^0$ 
intermediate state, which we will use as working example. A more careful 
analysis   
reveals that some of the above mentioned two-body states can be excluded because 
their interactions yielding a  $K^+\pi^+\pi^-$ final state would violate $G$ 
parity 
(this is the case of the $K^0 \pi^+$ and  $a_0^+ K^0$). Some other states are 
much 
less abundant as two body decays. This is the case of the $\omega K^+$ and 
$\omega K^{*+}$ states, which branching fraction is one order of magnitude 
smaller than 
the $K^+ \pi^0$ one. Other diagrams are suppressed by 
dual topological unitarity (DTU)  \cite{dtu}. Finally,  
some other states, inspite of being different, are expected to 
yield amplitudes which are qualitatively very similar. This is, for example, 
the case of the 
states  $K^+ \pi^0$ and $\eta' K^+$. Replacing $\pi^0$ by $\eta'$ in Fig. 
\ref{fsi1} will lead 
to a similar diagram with  the vertex $\eta' \, \rho \, \pi$ instead of 
$\pi \,  \rho \, \pi$. 
This will change the corresponding coupling constant and the form factor. The 
former is unable to distort 
the Dalitz plot and irrelevant for our discussion. In both vertices the form 
factor will 
suppress highly virtual internal $\rho$ mesons. We therefore expect these two 
diagrams to give 
similar results.  Even after a careful scrutiny and the elimination of 
suppressed diagrams and 
of ``double counting'', there will remain a few diagrams to be considered. 
Here we shall 
work out in detail only  the process depicted in  Fig. \ref{fsi1}. 
This will  show schematically how the 
final state meson dynamics will lead to inhomogeneities in the Dalitz plot.

Before starting  the evaluation of the  Feynman diagram  given in 
Fig. \ref{fsi1}, one last remark is in order. In other FSI calculations, such 
as \cite{china} 
the starting point is the $B$ meson, which decays into two off-shell  mesons, 
which, in 
turn exchange a third virtual meson. This process is described by a loop 
diagram, from which 
the absorptive part is considered. In our Fig. \ref{fsi1} 
this procedure would be equivalent to close the
lower part of the diagram forming a loop with a virtual $K^+$ and a virtual 
$\pi^0$. Here, for
simplicity,  we take them to be real. In this way we will give emphasis to the 
creation of the  
soft  $\pi^-$ and consequent Dalitz plot distortion. The loop integral 
will introduce 
some smearing in  our result and presumably change the normalization. Since our 
purpose in 
this work  is to discuss this process  qualitatively, we postpone to the future 
a more general calculation, including effects of the kaon and pion 
off-shellness.

The effective Lagrangians relevant to the calculation can be constructed 
from a chiral Lagrangian \cite{ko}. They are:

\begin{eqnarray}  \label{intlag}
\mathcal{L}_{\omega KK}&=& ig_{\omega KK} (\bar{K}\partial_\mu K-
\partial_\mu \bar{K}K){\omega}^\mu,  \nonumber \\
\mathcal{L}_{\rho \pi\pi}&=& g_{\rho\pi\pi} \vec{\rho}^\mu\cdot(\vec{\pi}
\times\partial_\mu \vec{\pi}),  \nonumber \\
\mathcal{L}_{\rho\omega\pi}&=&g_{\rho\omega\pi}\epsilon^{\al\be\la\si}
\partial_\be\omega_\al\vec{\rho}_\la\cdot\partial_\si\vec{\pi}.
\end{eqnarray}

With the above interaction Lagrangians, we can write the amplitude for the 
diagram in Fig. \ref{fsi1} as:
\beq
{\cal A}={4ig_{\omega KK}g_{\rho\pi\pi}g_{\rho\omega\pi}\epsilon^{\al\be\la
\si}p_{5\al}p_{2\be}p_{1\la}p_{3\si}\over\left((p_1-p_3)^2-m_\omega^2\right)
\left((p_5-p_2)^2-m_\rho^2\right)},
\label{amp}
\enq
where $p_1$ and $p_2$ denote  the momenta of the $K^+$ and
$\pi^0$ in the initial state; $p_3,~p_4$ and $p_5$ are 
those of the $K^+,~\pi^-$ and $\pi^+$ in the final state of the  
diagram in Fig. \ref{fsi1}. It is important to notice that $p_1+p_2=p_B$, where 
$p_B$
is the momentum of the $B$ meson, since we are considering the final state 
interactions in the two-body decay $B^+\to K^+\pi^0$. 

In order to take into account the effects of  hadron internal structure we follow 
ref.~\cite{ko2} and introduce, in the amplitude, the  form factor:
\beq
F(q_\omega,q_\rho)=\left(\Lambda^4\over\Lambda^4+(q_\omega^2-m_\omega^2)^2
\right)\left(\Lambda^4\over\Lambda^4+(q_\rho^2-m_\rho^2)^2\right),
\enq
where $q_\omega$ and $q_\rho$ are the four momenta of the intermediate 
off-shell vector mesons, {\it i.e.}, $q_\omega=p_1-p_3$ and $q_\rho=p_5-p_2$ 
and $\Lambda$ is a cut-off parameter taken to be $\Lambda = 1$ GeV. 

Parametrizing the
amplitude of the weak decay $B^+\to K^+\pi^0$ through the weak decay coupling
$G_{BK\pi}$, we can write the amplitude for the three-body decay,
$B^+\to K^+\pi^+\pi^-$,  in terms of the $K^+\pi^-$ ($\pkpi=p_3+p_4$)
and $\pi^+\pi^-$ ($\ppipi=p_4+p_5$) momenta as:
\beqa
&&{\cal A}_{K\pi\pi}=
{-32i\Lambda^8G_{BK\pi}g_{\omega KK}g_{\rho\pi\pi}g_{\rho\omega\pi}\over
\left(q_1^2-4m_\omega^2\right)\left(\Lambda^4+(q_1^2/4-m_\omega^2)^2
\right)}
\nn\\
&\times&{\epsilon^{\al\be\la\si}(\ppipi)_{\al}p_{B\be}P_{\la}(\pkpi)_{\si}
\over\left(q_2^2-4m_\rho^2\right)\left(\Lambda^4+(q_2^2/4-m_\rho^2)^2
\right)},
\label{ampf}
\enqa
where $q_1= P-p_B+2\ppipi$, $q_2=P+p_B-2\pkpi$ and $P=p_1-p_2$. 
In terms of these momenta, the three-body decay rate
is given by:
\beqa
d\Gamma={\left|{\cal A}_{K\pi\pi}\right|^2\over2^4\pi^5m_B}\delta\left(
(\ppipi+\pkpi-p_B)^2-m_\pi^2\right)\delta\left((p_B\right.
\nn\\
-\left.\pkpi)^2-m_\pi^2\right)\delta\left((p_B-\ppipi)^2-m_K^2\right)
~d^4\ppipi~d^4\pkpi.
\label{rate}
\enqa

Evaluating Eq.~(\ref{rate}) and using  the delta functions  
to perform some  of the integrals, we finally  write the decay rate in the
standard form for the Dalitz plot:
\beq
{d^2\Gamma\over ds_{13}ds_{23}}=\int{\left|{\cal A}_{K\pi\pi}\right|^2\over
2^8\pi^4m_B^2}~d(\cos\theta_{K\pi})~d\phi_{\pi\pi},
\label{frate}
\enq
where, as in Eq.~(\ref{anr}), $s_{13}=\pkpi^2$ and $s_{23}=\ppipi^2$.
In the rest frame of the $B$ meson and due to the delta functions, in 
evaluating $\left|{\cal A}_{K\pi\pi}\right|^2$ one has to use:
\beq
P^2=2m_\pi^2+2m_K^2-m_B^2,
\enq
 with 
\beq
E_P={m_K^2-m_\pi^2\over m_B} \mbox{ and }
|\vec{P}|
={\sqrt{\lambda(m_B^2,m_K^2,m_\pi^2)}\over m_B}, 
\enq
where $\lambda(x,y,z)=x^2+y^2+z^2-2xy-2xz-2yz$. For the other momenta and
scalar products we have:
\beq
q_1^2=2s_{23}-2m_B^2+2m_K^2+4m_\pi^2+4P.\ppipi,
\enq
\beq
q_2^2=2s_{13}-2m_B^2+2m_\pi^2+4m_K^2-4P.\pkpi,
\enq
\beq
P.\pkpi=E_PE_{K\pi}-|\vec{P}||\vec{p}_{K\pi}|\cos{\theta_{K\pi}},
\enq
\beqa
P.\ppipi&=&E_PE_{\pi\pi}-|\vec{P}||\vec{p}_{\pi\pi}|(\cos{\theta_{K\pi}}
\cos\bar{\theta}\nn\\
&+&\sin{\theta_{K\pi}}\sin\bar{\theta}\cos{\phi_{\pi\pi}}),
\enqa
\beqa
\cos{\bar{\theta}}&=&(m_K^2-m_\pi^2-m_B^2+s_{13}+s_{23}+
\nn\\
&+&\left.{(s_{13}-m_\pi^2)(s_{23}-m_K^2)\over m_B^2}\right)
{1\over4|\vec{p}_{K\pi}||\vec{p}_{\pi\pi}|},
\enqa
\beq
E_{K\pi}={m_B^2+s_{13}-m_\pi^2\over2m_B},
\enq
\beq 
E_{\pi\pi}={m_B^2+s_{23}-m_K^2\over2m_B},
\enq
\beq
|\vec{p}_{\pi\pi}|={\sqrt{\lambda(m_B^2,s_{23},m_K^2)}\over2m_B},
\enq
\beq
|\vec{p}_{K\pi}|={\sqrt{\lambda(m_B^2,s_{13},m_\pi^2)}\over2m_B},
\enq
\beq
\ppipi.\pkpi={m_B^2-m_K^2\over2}.
\enq 

\begin{figure}[h] 
\centerline{\epsfig{figure=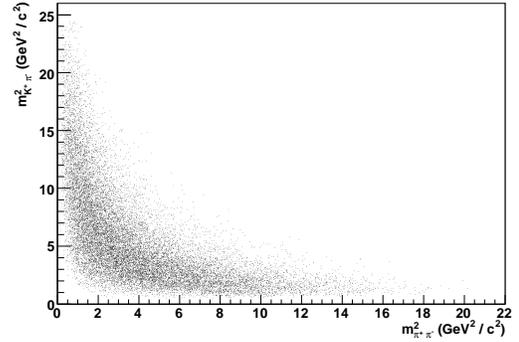,height=50mm}}
\caption{The Dalitz plot for the non-resonant $B^+\to K^+\pi^+\pi^-$ decay.}
\label{dali} 
\end{figure} 
For a given value of $s_{13}$ ($(m_\pi+m_K)^2\leq s_{13}\leq(m_B-
m_\pi)^2$), the 
range of $s_{23}$ is determined by imposing $-1\leq\cos{\bar{\theta}}\leq1$ 
and is given by \cite{pdg}:
\beqa
({s_{23}})_{min}&=&(E^*_1+E^*_2)^2-\left(\sqrt{E^{*2}_1-m_\pi^2}+\sqrt{
E^{*2}_2-m_\pi^2}\right)^2
\nn\\
({s_{23}})_{max}&=&(E^*_1+E^*_2)^2-\left(\sqrt{E^{*2}_1-m_\pi^2}-\sqrt{
E^{*2}_2-m_\pi^2}\right)^2
\nn
\enqa
where
\beqa
E^*_1&=&{s_{13}-m_K^2+m_\pi^2\over 2\pkpi},
\nn\\
E^*_2&=&{m_B^2-s_{13}-m_\pi^2\over 2\pkpi}.
\nn
\enqa

Before presenting our numerical results we would like to emphasize that 
{\it we do not wish to reproduce the absolute normalization of the decay 
rate}. The only purpose of our calculation is to show that the meson exchange 
mechanism is able to produce distortions in the Dalitz plot of the 
$B^+\to K^+\pi^+\pi^-$ decay.

In Fig.~\ref{dali} we show the Dalitz plot of a Monte Carlo simulation of 
Eq.~(\ref{frate}). The distribution shown 
in the figure is consistent  with the parametrization in Eq.(\ref{anr}) but it  
goes to zero near the  threshold. This result 
gives support to  the expectation presented in section II: 
the $K^+$ and $\pi^+$ carry the valence quarks from the intermediate $K^+$ and 
$\pi^0$  
and are  therefore hard, while the $\pi^-$ is created from the vacuum and is 
therefore softer.

The Dalitz distribution of Fig. \ref{dali} can be parametrized as 
\beq
|{\cal A}_{nr}(K^+\pi^-\pi^+)|^2 \, \propto  \,  \sqrt{s_{23} \, s_{13} } \, 
f_1(s_{23}) \, f_2 (s_{13}) \,
e^{-D s_{23}^2 s_{13}^2 }
\label{paradali}
\enq
where
\beq
f_i (x) \, = \, \frac{1}{1 \, + \, e^{\,[ \, c_i \,(x - p_i) ] } }
\label{fermi}
\enq
with  $D=1.3232 \times 10^{-3}$ GeV$^{-8}$, 
$c_1= 0.65$ GeV$^{-2}$, $p_1=18$ GeV$^2$, $c_2=0.55$ GeV$^{-2}$ and $p_2=15$ 
GeV$^2$.

%\section{Other three-body decays}

\begin{figure}[h] 
\centerline{\epsfig{figure=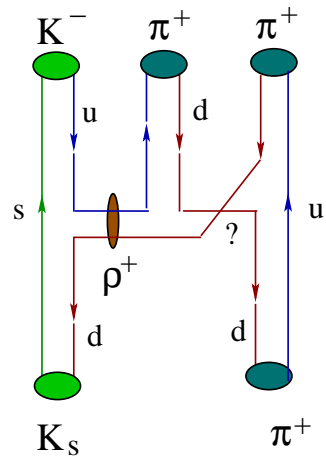,height=60mm}}
\caption{Born diagram for $K_s\pi^+\to K^-\pi^+\pi^+$ final state 
hadronic interaction.}
\label{diagra2} 
\end{figure}

\begin{figure}[h] 
\centerline{\epsfig{figure=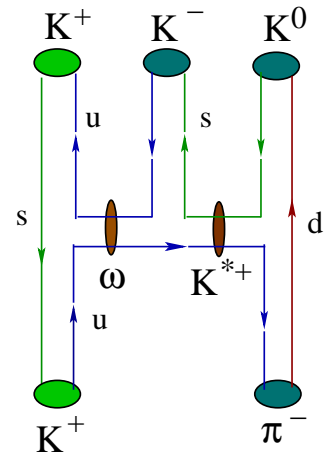,height=60mm}}
\caption{Born diagram for $K^+\pi^-\to K^+K^-K^0$ final state hadronic
interaction.}
\label{diagra3} 
\end{figure}

The meson exchange mechanism  considered above 
could play the same role in the $D^+\to K^-\pi^+\pi^+$ three-body decay, where 
one does not
see any inhomogeneity in the non-resonant amplitude. In the case of the
$D^+$, the three-body decay could proceed through final state interactions
between the $K_s$ and $\pi^+$ mesons in the two-body decay $D^+\to K_s
\pi^+$. In Fig.~\ref{diagra2} we show the Feynman diagram for the
$D^+\to K^-\pi^+\pi^+$ decay through the   $K_s \, \pi^+$ interaction. 
We see that there are no mesons that could be exchanged 
satisfying  all the conservation laws required by strong interactions. 
This means that the diagram does not exist and the three-body decay 
$D^+\to K^-\pi^+\pi^+$ does not
proceed through final state interactions. 
A similar situation occurs in the
thre-body decay $B^+\to D^-\pi^+\pi^+$, since there is no possible meson
exchange that would lead the two mesons $\bar{D^0}$ and $\pi^+$ in the
two-body decay $B^+\to \bar{D^0}\pi^+$ into the final three mesons:
$D^-\pi^+\pi^+$.   
On the other hand, the
three-body decay $B^0\to K^+K^-K^0$  can proceed through the  
two-body decay $B^0\to K^+\pi^-$, followed by final state  
interactions between  $K^+$ and $\pi^-$, 
as can be seen by the Feynman diagrams in Fig.~\ref{diagra3}.

Therefore, in the case of the three-body decay $B^0\to K^+K^-K^0$, one
expects some inhomogeneity in the non-resonant amplitude, consistent with a
hard $K^+$, a hard $K^0$ and a  softer $K^-$. This effect was not included in the 
data analysis performed by the BaBar Colaboration in ref.~\cite{babar1}, 
where a homogeneous parametrization for  the  non-resonant component was used.

To close this section we would like to comment the Belle results on baryonic 
three body decays. In this context  the FSI 
approach was first  used in \cite{hou} to study  rare baryonic $B$ decays.

The study of the decays $B^+ \rightarrow p \, \bar{p}\, K^+$, 
 $B^+ \rightarrow p \, \bar{p}\, \pi^+$ \cite{wang} revealed a low mass 
enhancement 
of the system $p - \bar{p}$. However this enhancement in the low mass (close to 
the threshold) baryon-antibaryon system  was not seen in the decays
$B \rightarrow p \, \bar{p}\, J/\psi$ and 
$B^- \rightarrow \Lambda \, \bar{p}\, J/\psi$ \cite{xie}. 
In terms of FSI we can very easily 
understand these results. The mass of the  baryon-antibaryon system is low 
because  the  $\bar{p}$ is soft.
In the case of light meson production  we have the 
following sequence of decays: 
$B^+ \rightarrow K^+ \, \pi^0 \rightarrow K^+ \, p \, \bar{p} $. The intermediate 
pion 
splits into a proton and an antiproton and the latter interacts with the kaon 
exchanging a $\rho$ or an $\omega$. The emerging $\bar{p}$ has all the antiquarks 
coming from the vacuum and will therefore carry low momentum. In the  
case of $J/\psi$ production the $B$ goes first to an intermediate state with
 $J/\psi$, such as, for example, 
$J/\psi \pi^0$. The pion could then split into $p$ and $\bar{p}$ but now the 
subsequent elastic interaction between the $J/\psi$ and the $\bar{p}$ is 
OZI suppressed, since the two mesons have  no quarks in common. Therefore, in this 
case there is no soft $\bar{p}$ from FSI and no low mass enhancement in the 
$p \bar{p}$ system. This same problem has been addressed in \cite{la} where a 
similar conclusion was obtained.

\section{Concluding remarks}

We hope to have made clear that, for experimental purposes, it is highly desirable 
to have a parametrization for the amplitude of non-resonant three body decays. 
This 
parametrization should have a physical basis and should be derived from  theory. 
The 
already existing parametrization given in Eq. (\ref{anr}) 
is not satisfactory from this point of view. 
In order to improve it  we have considered one possible mechanism responsible for 
inhomogeneities in the Dalitz plot, which we have called final state interactions (FSI).
We have sketched calculations of the Dalitz distributions  with this mechanism. These 
calculations can certainly be improved, but already at this stage they can teach us how 
to obtain distortions 
in the non-resonant three body decays. FSI   trace back the 
observed Dalitz distributions to a manifestation of the leading particle effect.

\section*{Acknowledgements}
{This work has been partly supported by FAPESP and CNPq.}

%%%%%%%%%%%%%%%%%%%%%%%%%%%%%%%%%%%%%%%%%%%%%%%%%%%%%%%%%%%%%%%%%%%%%%%%%%%%%
%%%%%%%%%%%%%%%%%%%%%%%%%%%%%%%%%%%%%%%%%%%%%%%%%%%%%%%%%%%%%%%%%%%%%%%%%%%%%%
%%%%%%%%%%%

%%%%%%%%%%%%%%%%%%%%%%%%%%%%%%%%

\end{document}